\documentclass[conference,compsoc]{IEEEtran}
\ifCLASSOPTIONcompsoc
  \usepackage[nocompress]{cite}
\else
  \usepackage{cite}
\fi
\ifCLASSINFOpdf
\else
\fi

\usepackage{multirow}

\usepackage{amssymb,amsmath}
\usepackage{booktabs}
\usepackage{graphicx}
\usepackage{tikz}

\usepackage{xspace}
\usepackage{xurl}
\usepackage{colortbl}
\usepackage{adjustbox} 
\usepackage{caption} 
\usepackage{subcaption}
\usepackage{tabularx}
\usepackage{makecell}
\usepackage{nicefrac}
\usepackage{array}
\usepackage{hyperref}
\usepackage{cleveref}
\usepackage{enumitem}
\newcommand{\myparagraph}[1]{\vspace{5pt}\noindent\textbf{#1.}\hspace{5pt}}
\newcommand{\censhaper}{\textit{CenShaper}}

\newcommand{\proxyshaper}{\textit{ProxyShaper}}
\newcommand{\persona}{persona}
\newcommand{\profile}{profile}
\newcommand{\schedule}{traffic schedule}

\usepackage{pifont}
\newcommand{\fstep}[1]{\raisebox{-0.08ex}{\scalebox{1.15}{\ding{\the\numexpr171+#1\relax}}}}

\begin{document}
%
\title{\censhaper{}: Shaping Proxy Traffic for Censorship Circumvention}

\author{\IEEEauthorblockN{Shabbir Hussain}
    \IEEEauthorblockA{
        University of Michigan\\
        shabbirx@umich.edu
    }
    \and
    \IEEEauthorblockN{Aaron Ortwein}
    \IEEEauthorblockA{
        University of Michigan\\
        aortwein@umich.edu
    }
    \and
    \IEEEauthorblockN{Roya Ensafi}
    \IEEEauthorblockA{
        University of Michigan\\
        ensafi@umich.edu
    }}


%


\maketitle

\begin{abstract}

Modern commodity DPI systems are capable of analyzing packet sizes, directions, and timings to detect circumvention tools even when those tools use sophisticated obfuscation. The anti-censorship community has called for innovation in traffic shaping and built an ecosystem that facilitates the adoption of new shaping techniques. However, a fundamental gap is halting progress: \textit{what} should circumvention traffic look like? Past experience with traffic shaping has underscored that effective shaping must produce traffic that looks like legitimate Internet traffic while limiting the ability of censors to identify multiple proxies through shared fingerprints.

In this paper, we use our access to large-scale real-world TLS traffic at a regional ISP to analyze approximately 12 million connections, finding that individual servers differ substantially from the broader TLS population and that traffic patterns vary across services. Combining these empirical observations with past experience with traffic shaping, we build \censhaper{}, which shapes each proxy's traffic to look like a selected TLS server. To reproduce these server-specific patterns, \censhaper{} trains a shared generative model using traffic from TLS connections to many servers and adapts it for each proxy using a small sample of recorded connections to its selected server. Our model evaluation shows that fingerprints learned from known proxies transfer poorly to proxies imitating previously unseen TLS services. In end-to-end testing, \censhaper{} reduces the detection rate of TLS-over-TLS fingerprinting from over 85\% to approximately 3\% at a 1\% false positive rate. Finally, we show how running \censhaper{} as a sidecar process simplifies integration with existing circumvention tools, making it a practical defense against increasingly powerful traffic analysis attacks.
\end{abstract}

%
\IEEEpeerreviewmaketitle

\section{Introduction}
\label{sec:intro}

Deep packet inspection (DPI) devices have become increasingly capable of identifying censorship circumvention traffic through advanced analysis of packet sizes, directions, and timings. China and Russia have blocked a wide range of proxy protocols, and prior reports suggest that these campaigns relied on traffic patterns inherent to tunneling~\cite{chinaTLSinTLS,russiaTLSinTLS}. More recently, leaked code from Geedge Networks, a Chinese DPI company with ties to the Great Firewall, revealed the use of packet sizes and directions to fingerprint circumvention tools~\cite{Ablove2026a}. Geedge has exported its products to several countries, including Kazakhstan, Ethiopia, Pakistan, and Myanmar, lowering the barrier to deploying advanced traffic analysis globally. These developments demonstrate that advanced traffic-analysis attacks are no longer a distant possibility, but an increasingly pressing threat to censorship circumvention worldwide.

Luckily, much of the technical groundwork for defending against traffic analysis is already in place: recent work has shown that traffic shaping can operate as a plug-and-play layer on top of existing proxies rather than being built into each protocol~\cite{Ortwein2026a,Wilson2025a,Johnson26a}, and systems such as WATER~\cite{Chi2024a}, Outline SDK~\cite{outline}, and Proteus~\cite{Wails2023a} enable rapid distribution and deployment of new shaping strategies. What remains open is a more fundamental question: \textit{what} shape should proxy traffic take? Early solutions such as ScrambleSuit~\cite{Winter2013b} and obfs4~\cite{obfs4} randomize packet sizes and timings to give different proxy servers distinct traffic shapes, aiming to prevent censors from identifying them all using a single fingerprint. However, randomized traffic shapes stand out from ordinary Internet traffic, allowing censors to identify and block them~\cite{Wails2024a,Xu2021,Huang2026}. 

Another approach is to make proxy traffic mimic ordinary Internet traffic, so that censors cannot block proxies without also disrupting legitimate connections~\cite{Moghaddam2012a,Dyer2015a,Ortwein2026a}. Yet the anti-censorship community has long held that perfect mimicry is difficult~\cite{Houmansadr2013a}. Indeed, even proxies designed to look like ordinary TLS traffic can be told apart by patterns in packet sizes, directions, and timings~\cite{Ortwein2026a}, including patterns created by TLS handshakes inside the tunnel~\cite{Xue2024a} and by differences between transport- and application-layer round-trip times~\cite{Xue2025a}. 

Since neither randomization nor mimicry is enough on its own, an effective shaping strategy could combine the two, making proxies more difficult to distinguish from ordinary traffic while limiting the ability of censors to reuse fingerprints learned from individual proxies. What if a proxy provider shaped \textit{each} proxy to imitate a distinct legitimate TLS server? We call the TLS server a proxy imitates its \emph{persona}. By assigning each proxy a different \persona{}, such as a particular website or CDN, the provider would end up with proxies that collectively exhibit many distinct traffic shapes. This diversity would make it harder for a censor to learn a single fingerprint that blocks the provider as a whole.

In this paper, we investigate whether this \emph{ordinary, but not identical} approach is a good defense against traffic analysis. Whether this idea works depends on two sets of questions. The first concerns properties of real-world traffic that are difficult for the community to measure: Do different TLS servers have distinct enough traffic shapes to give proxies meaningful diversity? Do these shapes stay stable over time? And how much traffic must a proxy provider record to represent a \persona{}? The second concerns whether the idea can be built: Can a shaper efficiently learn and generate a server's traffic shape? Can a client and proxy independently derive the same shape, avoiding the synchronization problems that have long complicated censorship circumvention? And can all of this be packaged as a plug-and-play layer that providers can deploy as one more strategy in their defenses? Answering the first set requires observing many real servers, at scale and over time, from the network's point of view, a vantage point that proxy providers typically lack. Answering the second requires building and evaluating such a shaper.

To answer the first set of questions, we collaborate with a regional ISP that serves
a population of 1 million users to study real-world TLS traffic. In \Cref{sec:analysis}, we show that traffic shapes differ substantially across TLS servers, while servers hosting the same domain exhibit similar shapes across IP addresses. In other words, legitimate traffic offers both a well-defined target for each proxy server and meaningful diversity across the proxy servers. We also find that these shapes remain relatively stable over three days, and that for most busy servers, only 160 recorded flows are enough to represent a server. Together, these measurements provide the empirical foundation on which the \emph{ordinary, but not identical} approach rests.

To answer the second set of questions, we build \censhaper{}, a traffic shaper that makes proxy connections look like ordinary connections to a selected \persona{}. At its core, \censhaper{} uses generative decision forests to learn a general model of TLS traffic from many servers, capturing dependencies among packet sizes, directions, timings, and TLS records. Each proxy then adapts this model to its \persona{} using a small number of the \persona{}'s recorded flows, without retraining the forests. This reduces the load on the provider, who only needs to record connections to a \persona{}, not train a new model. For each proxy connection, the adapted model generates a \emph{\schedule{}}: the direction, size, delay, and TLS records of each of the first 20 data packets, as a real connection to the \persona{} might produce them. \censhaper{} then makes the proxy connection follow the shape of the \schedule{},  padding, splitting, and delaying packets as needed. Because the forests are deterministic, the client and proxy can each generate the same synthetic flow from their shared TLS handshake secret. Doing so removes the need for synchronization messages, which could themselves become fingerprints. Lastly, \censhaper{} runs independently of the proxy and handles its outer TLS connection, allowing it to be integrated with existing proxies without code changes. We implement \censhaper{} as a WATER transport module, allowing shaping strategies to be updated without any redeployment.

We evaluate \censhaper{} against classifiers from prior work and known traffic-analysis attacks. First, can a censor that has learned the shape of one proxy server use it to identify others? We train classifiers from prior work (Deep Fingerprinting, p-FP, Stacked Denoising Autoencoder, and XGBoost) on the synthetic flow shapes our generator produces and set their false positive rate (collateral damage) to a generous 1\%. The best-performing classifier detects only 8.5\% of traffic shapes generated using \persona{}s drawn from previously unseen TLS servers. When the classifier has seen traffic shapes generated using other servers of the same TLS service, detection rises to 29\%. When it has seen traffic shapes from the evaluated persona itself, detection reaches 56\%. These results suggest that \persona{} diversity limits what a censor can reuse across proxies. We then build a testbed and use \censhaper{} to shape the traffic of five popular proxy protocols---VLESS, VMess, VLESS+WS, Trojan, and Shadowsocks. Using the same classifier, a censor detects only 8.2\% of flows from unseen TLS servers. However, when the classifier has seen training examples generated using every \persona{}, it detects 77\% of proxy traffic. Finally, we find that \censhaper{} makes proxy traffic harder to distinguish from ordinary TLS: the detection rate of the TLS-over-TLS fingerprinting attack~\cite{Xue2024a} drops from over 85\% to about 3\%, and TLS Divergence~\cite{Ortwein2026a} drops by roughly sevenfold. 

Traffic-analysis attacks are arriving faster than circumvention tools can prepare for them, and anti-censorship tool developers rarely have the resources to explore defenses before they are needed. This paper aims to fill that gap for traffic shaping. We show that legitimate TLS traffic can give each proxy a realistic and distinct shape, build \censhaper{} to put the \emph{ordinary, but not identical} approach into practice, and outline how providers could deploy it. Circumvention developers have kept their tools reachable in censored regions through some of the most aggressive blocking campaigns, adapting to active probing, IP blocking, and protocol fingerprinting as each emerged. We believe \censhaper{} can become one more tool in that defense, ready before traffic analysis becomes the next widespread threat. 
\section{Background and Related Work}
\label{sec:background}

Even when their contents are encrypted, proxy connections can exhibit distinctive patterns in packet sizes, directions, and timings. For example, TLS handshakes carried inside proxy tunnels can produce recognizable packet-size and direction patterns~\cite{Xue2024a}, while proxies can introduce larger discrepancies between transport- and application-layer round-trip times by acknowledging client packets before receiving responses from their destinations~\cite{Xue2025a}. Traffic shaping seeks to conceal these patterns~\cite{Wilson2025a,Johnson26a}, but systems differ in how they determine the resulting traffic shape.

\myparagraph{Random Shape} Several proxies, including ScrambleSuit~\cite{Winter2013b}, obfs4~\cite{obfs4}, VMess, and XTLS-Vision~\cite{Xue2024a}, use randomized padding to obscure traffic patterns. ScrambleSuit and obfs4 can also introduce randomized delays. However, randomness can itself provide a fingerprint: Wu et al.\ showed that the Great Firewall can identify fully encrypted proxies from their random-looking payloads~\cite{Wu2023a}. Randomizing packet sizes and timing also does not eliminate protocol-agnostic fingerprints inherent to tunneling. Xue et al.\ demonstrated that packet ordering and direction patterns can expose encapsulated TLS handshakes despite random padding~\cite{Xue2024a}. Similarly, obfs4's random delays can exacerbate discrepancies between transport- and application-layer round-trip times~\cite{Xue2025a}.

\myparagraph{Adversarial Perturbation}
Adversarial machine learning approaches use information from traffic classifiers to find perturbations to packet sizes and timings that induce misclassification while minimizing changes to the original traffic~\cite{Verma2018a,Granados2020a,Liu2022a,Liu2023a}. Although these perturbations can transfer across classifiers~\cite{Tramer2017a,Liu2023a}, retraining a classifier on adversarial traffic may render them ineffective~\cite{Liu2023a}. Moreover, inducing misclassification does not ensure that the resulting traffic looks like legitimate connections.

\myparagraph{Application Mimicry}
Another option is to give the tunnel the shape of an application the censor allows, on the premise that the application is too costly (economically or socially) to block. SkypeMorph shapes packet sizes and timing to match Skype video calls~\cite{Moghaddam2012a}, CensorSpoofer imitates voice-over-IP traffic~\cite{Wang2012a}, and StegoTorus replays the packet sizes and timing of recorded voice calls and imitates HTTP with recorded requests and responses~\cite{Weinberg2012a}. Marionette also models HTTP traffic, reproducing distributions of response lengths and connection counts measured from visits to a particular website~\cite{Dyer2015a}. Faithfully imitating an application's traffic, however, can be difficult without running the application itself~\cite{Houmansadr2013a}.

Some systems instead tunnel data through genuine applications while attempting to preserve their normal traffic characteristics. Facet~\cite{Li2014a} uses Skype video calls to carry censored videos, DeltaShaper~\cite{Barradas2017a} uses them to carry arbitrary data, and CovertCast~\cite{McPherson2016a} broadcasts encoded web content through live-streaming services. Protozoa~\cite{Barradas2020a} replaces encoded video-frame data in WebRTC calls, and Balboa~\cite{Rosen2021a} rewrites application traffic while preserving record lengths. However, these approaches require operating the underlying cover applications, limiting their ability to serve as plug-and-play traffic shapers for existing proxy protocols.

\myparagraph{TLS Mimicry} Rather than imitate a particular application, a proxy could shape toward the broader distribution of TLS traffic, taking advantage of its ubiquity and diversity. Ortwein et al.\ introduced CenKL to model TLS traffic and used it in ProxyShaper to shape the first ten post-handshake TLS records of VLESS and VMess-over-TLS, making both protocols harder for machine learning classifiers to distinguish from ordinary TLS~\cite{Ortwein2026a}. However, ProxyShaper uses the same population-level model for every proxy, raising questions about its consistency with individual TLS servers and whether fingerprints learned from one proxy could identify others. We investigate these questions by comparing individual servers with the TLS population and evaluating fingerprint transfer across distinct service-specific shaping targets (\Cref{sec:analysis,sec:eval}).

\begin{figure}[t!]
\centering
 \includegraphics[width=\columnwidth,keepaspectratio]{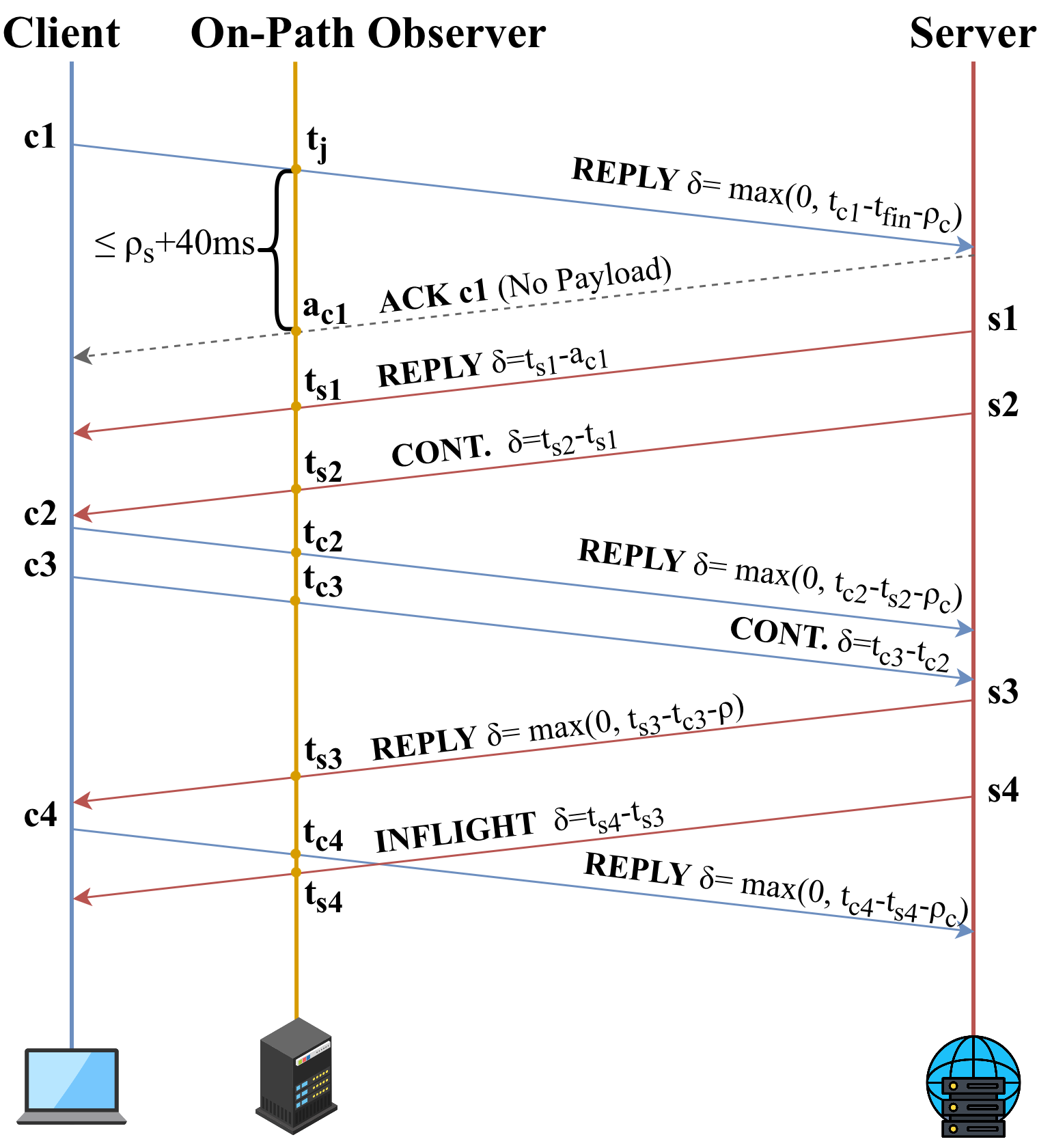}
\caption{\textbf{Delay Estimation.}
Packet labels show type and estimated delay $\delta$.
Replies are measured from a prompt ACK (s1) or from the most recently acknowledged packet, accounting for network travel time (s3). Continuations use the preceding packet (s2), while in-flight
packets use the preceding same-sender packet (s4).
The first packet (c1) is referenced to the TLS handshake.}
\label{fig:packet-type}
\end{figure}

\section{Threat Model and Data}
\label{sec:setting}

This section establishes our threat model and describes the TLS traffic used throughout the paper. We outline the censor's capabilities, then explain how we collect and represent traffic. We consider an on-path censor that observes packet sizes, directions, arrival times, and TLS record lengths but cannot decrypt TLS traffic. The censor can become a client of known proxies to learn their traffic patterns and identify others whose addresses are unknown. Because proxy traffic is rare relative to ordinary traffic, detection must operate at a low false positive rate. We assume proxies use uTLS~\cite{utls} to mimic a browser's TLS \texttt{ClientHello}. Our analysis focuses on the initial post-handshake packets, consistent with prior observations of early blocking decisions. Active probing of suspected proxies and detection beyond the shaped window are outside our scope.

\subsection{Data Collection}

We collected network traffic at a major point of presence of a regional U.S. ISP, where it was mirrored to an on-site server. We used Zeek to parse TLS traffic on TCP port 443 and assumed that most of the traffic identified by Zeek corresponds to regular Internet traffic rather than proxy traffic~\cite{Xue2024a}. For each packet, we logged its direction, TCP payload size, timestamp, TCP sequence and acknowledgment numbers, and number and lengths of TLS records beginning in it. For each flow, we logged the destination IP address and SNI. Since a single IP address can host multiple domains (e.g., in the case of CDNs), we define a server by its destination IP address and hostname, compared case-insensitively and a \emph{service} as all servers sharing a hostname. In our dataset, 99.8\% of TLS 1.3 flows and 97.7\% of TLS 1.2 flows specify the SNI. No packet payloads were logged.

We analyze the first 20 post-handshake data packets of each flow, excluding flows with fewer packets or with TCP sequence number gaps, retransmissions, or unknown TLS record counts within this window. Censors have identified and blocked circumvention tools using only the first data-carrying packet~\cite{Ablove2026a,Wu2023a}, making our analysis window substantially larger than what has been required in practice.

\subsection{Traffic Representation}
\label{sec:traffic-repr}

We represent each packet using features previously used for traffic identification: direction and TCP payload size~\cite{Wails2024a,Xue2024a}, timing~\cite{Xue2025a}, and the number of TLS records beginning in it~\cite{Oh2019a}. Specifically, we represent packet $i$ as
\[
    z_i = (d_i, s_i, c_i, b_i),
\]
where $d_i$ is its direction, $s_i$ its TCP payload size, $c_i$ the number of TLS records beginning in it, and $b_i$ its estimated delay bucket, defined below. We represent each flow by concatenating its packet feature vectors, $X = (z_1, \ldots, z_{20})$. We standardize each feature to zero mean and unit variance across all connections to prevent packet sizes from dominating the distance calculations used in our analysis~\cite{Hastie2009a}.

\myparagraph{Packet Order} 
\label{sec:packet-order}
When the client and server send packets at nearly the same time, their observed order can depend on the observer's location. We reconstruct the order as seen by the client, since censors observe traffic close to their users: for 70\% of the endpoints behind Russia's TSPU devices, the device is at most two hops away~\cite{Xue2022a}. 
Each client packet's TCP acknowledgment number indicates how much server data the client has received in order. We therefore place each client packet after the last server packet it acknowledges and before any server packets it has not yet acknowledged.

\myparagraph{Internal Response Time Estimation}
\label{sec:irt-est}
At an on-path observer, such as an ISP, the time between packets can reflect both sending delays and network propagation delay. However, a traffic shaper can only control when packets are sent, not how long they take to traverse the network. We therefore estimate each sending delay by measuring the time difference relative to an appropriate \emph{reference} packet.

Let $t_i$ be the time packet~$i$ passed the tap. If the preceding packet came from the same sender, we simply measure the time between the two packets. Otherwise, we use TCP acknowledgments (ACKs) to identify incoming traffic that the sender could be responding to. Let $j$ be the latest packet acknowledged by the sender before sending packet~$i$. A prompt, standalone ACK of $j$ provides a reference close to when the sender received $j$, allowing us to estimate the delay before sending packet~$i$. We consider an ACK prompt if it passed the tap within $\rho + 40$\,ms of $j$, where $\rho$ is the shortest round trip between the tap and the sender observed in the connection. When such an ACK is available, we measure from its observed time $a_j$. The 40\,ms threshold corresponds to Linux's minimum delayed-ACK time~\cite{linuxtcph}, the shortest among Windows, macOS, and Linux. In our dataset, delayed ACKs also cluster around 40 ms beyond the observed round-trip time. If the sender has not acknowledged any new packets since its previous packet~$p$, or if the ACK for $j$ was not captured, we measure from $p$. In the latter case, the estimate may include time before the sender received $j$. Otherwise, we measure from $j$ and subtract $\rho$ to approximate propagation delay, clipping negative estimates to zero:

\begin{equation*}
\delta_i =
\begin{cases}
t_i - t_{i-1}, & \text{packet $i-1$ from same side},\\
t_i - a_j, & \text{prompt ACK of packet $j$},\\
t_i - t_p, & \text{no new ACKs since $p$},\\
\max(0, t_i - t_j - \rho), & \text{otherwise.}
\end{cases}
\end{equation*}

\noindent We refer to a packet that directly follows one from the same sender as a \emph{continuation}. A packet that follows packets from the other sender is a \emph{reply} if its sender had acknowledged at least one of them, and \emph{in flight} if it had acknowledged none. \Cref{fig:packet-type} illustrates each case for server packets. 

Because these delays are estimates, we bucket them rather than retain exact values. Delays are also right-skewed~\cite{Ortwein2026a}, so we use base-2 buckets, $b_i = \lceil \log_2(\delta_i + 1) \rceil$, with $\delta_i$ in microseconds. This compresses long delays that would otherwise dominate the feature's scale.
\begin{figure*}[t]
  \centering
  \includegraphics[width=0.9\textwidth]{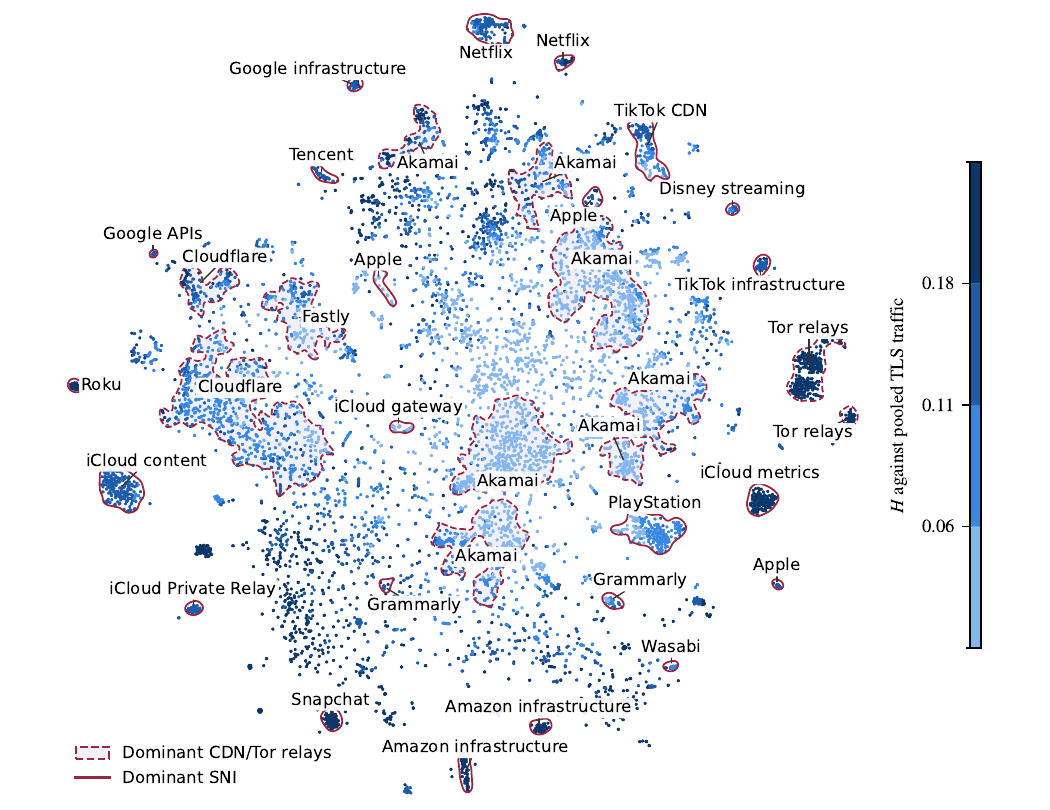}
  \caption{\textbf{Traffic shapes across destination addresses.} Each point is a destination address with at least 40 recorded TLS 1.2 and TLS 1.3 flows combined. We use t-SNE~\cite{Maaten2008a} to place addresses using the square root of their pairwise energy distance. Darker points have a larger energy coefficient \(H\) against the sampled population. Solid outlines mark selected areas dominated by an observed service name; dashed outlines mark selected operator-dominated areas or Tor relays.}
  \label{fig:map}
  \label{fig:map-distance}
\end{figure*}

\section{Empirical Analysis of TLS Across Servers} 
\label{sec:analysis}

TLS traffic observed at an ISP combines connections to many servers, each potentially exhibiting different traffic patterns. To understand the diversity of legitimate traffic available to our \textit{ordinary, but not identical} approach, we compare the traffic distributions of individual servers with the broader TLS population and with one another.

\subsection{Comparing Distributions}
\label{sec:comparing-dist}
To compare traffic distributions, we use \emph{energy distance}, denoted as $\mathcal{E}$~\cite{Szekely2013a}. Energy distance takes twice the average Euclidean distance \textit{across} the distributions and subtracts the average distance \textit{within} each. For connections $X, X'$ drawn independently from a distribution $P$ and $Y, Y'$ from $Q$:
\begin{equation}
  \mathcal{E}(P,Q)= 2\mathbb{E}\lVert X - Y\rVert - \mathbb{E}\lVert X - X'\rVert - \mathbb{E}\lVert Y - Y'\rVert
\end{equation}
Because $\mathcal{E}$ has no upper bound, we normalize it by twice the average distance between flows of the two populations:
\begin{equation}
H = \frac{\mathcal{E}(P,Q)}{2\,\mathbb{E}\lVert X - Y\rVert}
\end{equation}
The resulting energy coefficient~\cite{Fan2025a} $H$ ranges from 0 for identical distributions to at most 1. We report $H$ as a percentage to make small values easier to read. 

In order to reduce the influence of small sample sizes, we require a minimum of 40 flows per server which retains 90\% of traffic for both TLS~1.2 and TLS~1.3. Additionally, to limit the computational cost of calculating pairwise energy distance we consider at most 1,024 flows per server.

\myparagraph{Server Against Population}
We compare each server against a random sample of 2,000 flows from other addresses meeting the 40-flow minimum. The broader TLS population differs from nearly every server: the median server's $H$ against the population is 14\% for TLS 1.3 and 19\% for TLS 1.2, and only a tenth of servers have an $H$ below 7\%. 

We next examine how reliably connections to an individual server can be distinguished from other TLS traffic using only a few observed flows. To do so, we evaluate a simple classifier based on energy distance. For each server, we created two reference groups: half of the server's flows and the 2,000-flow sample. We then built a separate test set using the remaining server flows and an equal number of ISP flows, ensuring no overlap with the reference groups. Finally, we classified each test flow as either a ``server flow'' or ``population flow'' based on which reference group had a lower $H$ value. Based on this rule, the median classification accuracy for a single flow is 90\%. When classifying ten flows together, accuracy is above 95\% for 99.5\% of servers.

\myparagraph{Server Against Server}
The median energy coefficient between two randomly selected servers is 29\%, but drops to just 0.36\% for servers of the same service at different IP addresses. We visualize these similarities in \Cref{fig:map}, where addresses hosting the same service (e.g., Netflix and Snapchat) form distinct clusters near the edges. In contrast, addresses whose traffic most closely resembles the broader TLS population are predominantly CDNs. The median $H$ against the population is 5.9\% for Akamai, 8.6\% for Fastly, and 10.2\% for Cloudflare. Across these CDNs, greater similarity to the population corresponds to more domains hosted per IP address: Akamai hosts a median of 16 distinct domains, compared with 8 for Fastly and 3 for Cloudflare. 

\subsection{Recording Personas}
\label{sec:analysis-persona}
\begin{table}[t]
\centering\footnotesize
  
  \setlength{\tabcolsep}{4pt}
  \begin{tabular}{@{}l r r r r r@{}}
    \toprule
    & & & \multicolumn{3}{c}{Median $H$ after} \\
    \cmidrule(l){4-6}
    & Servers & Flows & 24 h & 48 h & 72 h \\
    \midrule
    TLS~1.3 & 3,516 & 5,310,115 & 0.12\% & 0.20\% & 0.19\% \\
    TLS~1.2 & 495 & 720,862 & 0.25\% & 0.32\% & 0.39\% \\
    \bottomrule
  \end{tabular}
  \caption{Median $H$ between flows of a server captured during the same four hour of the day on different days.}
  \label{tab:ttl}
\end{table}
\begin{figure*}[t]
  \centering
  \includegraphics[width=\textwidth]{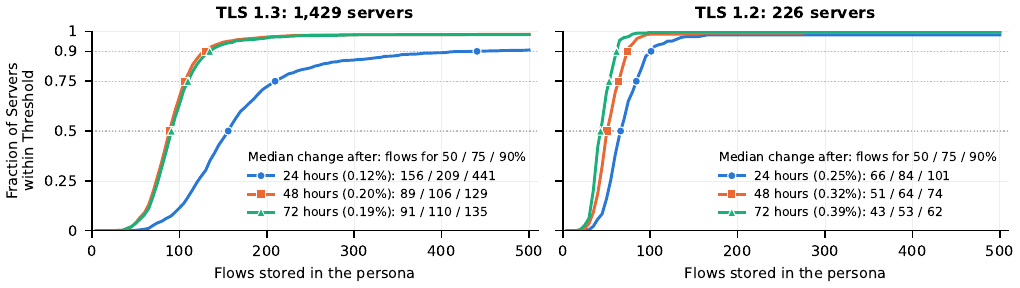}
  \caption{\textbf{Fraction of $n$ flow \persona{}s within threshold.} Evaluated over servers with at least 1,000 flows in our 24 hour capture and are present in all four windows. Each threshold is set based on servers change over time as reported in \Cref{tab:ttl}.
  }
  \label{fig:persona-cdf}
\end{figure*}

A proxy's \persona{} is the TLS server whose traffic it imitates. The proxy learns that traffic from a sample of the server's recorded flows, its \emph{\persona{} flows}, and imitates the server only as well as these flows represent its traffic. This raises two questions: how stable is a server's traffic distribution over time, and how many \persona{} flows are needed to represent it?

\myparagraph{Temporal Stability} To assess temporal stability, we compare traffic from the first four hours of capture with traffic from the same four-hour period 24, 48, and 72 hours later, considering only servers observed in all four periods. As shown in \Cref{tab:ttl}, the median $H$ ranges from 0.12\% to 0.20\% for TLS 1.3 and from 0.25\% to 0.39\% for TLS 1.2. These differences are small compared with the median $H$ of 29\% between randomly selected servers, suggesting that a server's traffic distribution remains relatively stable over several days.

\myparagraph{Number of Flows} We select each server's $n$ \persona{} flows greedily to approximate a server's traffic distribution. Starting with no flows, we repeatedly add the flow that minimizes the energy distance between the selected flows and the server's full set of recorded flows. To assess how closely $n$ \persona{} flows approximate its server's traffic distribution, we calculate $H$ between the \persona{} flows and at least $n$ other recorded flows from the same server, requiring at least $2n$ flows in total. We use the median temporal variation across servers as a benchmark: \persona{} flows should differ from its server's held-out flows by no more than the median change in a server's traffic over a set period of time (\Cref{tab:ttl}).

We test $n=5,10,15,\ldots,500$, restricting our analysis to servers with at least 1,000 flows in our 24-hour capture so that the same servers are compared across values of $n$. \Cref{fig:persona-cdf} shows the fraction of servers whose \persona{} flows meet each threshold as $n$ increases. Although the number of flows required varies across servers, 160 \persona{} flows meet the 24-hour threshold for at least 50\% of servers and the 48- and 72-hour thresholds for at least 90\% of servers, for both TLS versions. We therefore choose 160 flows per \persona{} for our model.

Restricting our analysis to servers with at least 1,000 flows favors busier servers. To examine this limitation, we repeat the analysis using all servers with at least $2n$ flows for each $n$, rather than imposing a fixed minimum of 1,000 flows. The results show similar trends, as shown in Appendix~\Cref{fig:appendix-persona-cdf}.

\subsection{Design Goals}
Our analysis shows that traffic patterns differ substantially across TLS services, that servers of the same service have similar patterns, and that a server's traffic can be represented by a small number of flows that remain representative for days. These findings motivate our \textit{ordinary, but not identical} approach: each proxy imitates a legitimate TLS service, while different proxies use distinct services. Based on this approach, we define three design goals for \censhaper{}:

\myparagraph{\textit{Goal 1}: Server Coherence}
A proxy's flows should follow the traffic distribution of the service it imitates, both individually and collectively (\Cref{sec:comparing-dist}). Switching between service-specific traffic patterns within a flow could expose an imitation, even if individual packets appear realistic~\cite{Houmansadr2013a}. Consistency also matters across flows: in our data, addresses with mixture-like traffic distributions tend to host many domains, such as CDNs, whereas flows to a single service exhibit similar patterns. Since a proxy presents a single SNI at its address, its flows should consistently follow that service's distribution.

\myparagraph{\textit{Goal 2}: Compartmentalization}
A censor can train a classifier on traffic to a known proxy (\textit{e.g.,} by becoming a client). Since servers hosting the same service have similar traffic distributions (\Cref{sec:analysis}), imperfections in imitation~\cite{Houmansadr2013a} could produce fingerprints transferable to other proxies imitating that service. Like polymorphic systems including ScrambleSuit~\cite{Winter2013b}, obfs4~\cite{obfs4}, and UPGen~\cite{Wails2025a}, we require that each proxy's traffic differs from every other proxy's to limit this transfer.

\myparagraph{\textit{Goal 3}: Limited Data}
Like previous traffic analysis studies~\cite{Ortwein2026a,Wails2024a,Xue2024a}, our analysis relies on a large-scale dataset collected through a partnership with an ISP. However, operators of circumvention services may not have access to such data, so \censhaper{} should not depend on it. We therefore require that \censhaper{} can learn realistic TLS traffic patterns from relatively little data. Our analysis suggests that this is achievable for an individual server, whose traffic can be represented by 160 flows (\Cref{sec:analysis-persona}).
\section{\censhaper{}}
\label{sec:censhaper}

In this section, we present \censhaper{}, a traffic shaper which sits between the endpoint's application layer and the proxy process running on it. We implement \censhaper{} with a \textit{generator} that produces server-specific traffic shapes and a shaper that transmits proxy traffic according to that shape.

\subsection{Generator}
\label{sec:model}

To satisfy G1, \censhaper{} must first generate traffic shapes that follow the distribution of a real TLS server. We call the component that produces these shapes the \emph{generator} and the generated shape a \textit{\schedule{}}. It has two parts: (1) a base model of generative decision forests, grown once from the flows of many servers and shared by every proxy, and (2) a \textit{\profile{}}, one per proxy, which makes the forests produce flows similar to that of its selected \persona{}. Although TLS servers have distinct traffic distributions (\Cref{sec:analysis}), their connections share structural constraints and recurring patterns. For example, TLS records can span multiple packets, while replies typically follow requests after longer delays than packets within a burst. The forests learn these relationships from thousands of flows, while each \profile{} supplies server-specific frequencies from just 160 \persona{} flows (G3).

\myparagraph{Why Decision Forests}
Recent generators of traffic shapes for evasion and defense use neural networks \cite{Gong2022a,Wails2022a,Liu2023a}. However, floating-point computations in neural networks can produce different results across hardware and software implementations, even with identical models and inputs~\cite{Schlogl2023a,pytorchnumerics}. Existing neural-network-based shapers instead use additional messages or shared state to coordinate their traffic shapes. For instance, Surakav's~\cite{Gong2022a} client generates the packet shape and tells the proxy exactly how many bytes to send back, while Raven~\cite{Wails2022a} sends parameters of its generator in the handshake. Amoeba~\cite{Liu2023a} generates flow profiles in advance and stores them in a database kept in sync at both ends. Each of these tools adds messages or shared state that a censor could observe or that must be distributed.

We instead use \emph{generative decision forests}~\cite{Bahl1989a} for two reasons: First, a tree generates a value by drawing from integer counts stored at its nodes, so a client and a proxy that start from the same random seed generate exactly the same \schedule{} without exchanging any messages. Second, these counts can be replaced without changing the trees, which lets us adapt the generator to a single server using a small number of its flows.

\myparagraph{Traffic Schedule} For comparing flows of real servers (\Cref{sec:analysis}) we only required the number of TLS records beginning in a packet and the bucket of its delay. However, a \censhaper{} also requires the length of each record and the exact delay from its \schedule{}. For a \schedule{}, we therefore describe each data-carrying TCP packet $x_i$ by:
\begin{equation*}
  x_i = (d_i,\; s_i, \; \ell_i, \; \delta_i).
\end{equation*}
Here, $d_i$ and $s_i$ are the packet's direction and TCP payload size as in \Cref{sec:traffic-repr} (sizes reach a maximum segment size of 1,460 bytes), $\ell_i$ the lengths of all TLS records beginning in the packet, and $\delta_i$ the delay relative to an earlier reference packet in milliseconds. The generator first draws the delay's base-2 bucket (\Cref{sec:irt-est}) and then a specific delay within it. To do so, we subdivide each bucket into equal-width intervals. The number of intervals follows Sturges' rule~\cite{Sturges1926a} for the $n$ training delays observed in that bucket, up to at most 32. The generator samples an interval according to its observed frequency, then selects a delay uniformly at random within that interval.

We also introduce two symbolic encodings for recurring patterns we observe in how TLS records align with packet boundaries. These reduce model complexity by allowing observations with different byte counts to contribute to the same learned pattern, rather than treating each size as a separate outcome. Approximately 17\% of packets that continue a TLS record from an earlier packet end with the record; we encode the sizes of these packets as \textsc{tail} rather than as absolute byte counts. Similarly, 65\% of TLS records span an entire packet, and their lengths are encoded as \textsc{fill}.

\myparagraph{Reference Packet}
A packet's delay is measured from a reference that depends on whether the packet is a continuation, a reply, or in-flight (\Cref{sec:irt-est}). To reproduce these timing relationships, our model includes each packet's type $k_i$ and reference $r_i$, expressed as the number of packets back from the current one. The first data packet instead uses the TLS handshake as its reference. Our full packet representation is therefore:

\begin{equation*}
  x_i = (d_i,\; s_i,\; \ell_i,\; \delta_i,\; k_i,\; r_i).
\end{equation*}

\myparagraph{Packet Generation}
Given sequence of packets generated so far, $h = (x_1, \ldots, x_{i-1})$, our generator calculates the probability of shape for next packet $x_i$ as:
\begin{multline*}
  P(x_i \mid h) = P(d_i \mid h)\; P(s_i \mid h, d_i)\;
  P(k_i, r_i \mid h, d_i, s_i)\\ P(\ell_i \mid h, d_i, s_i)\;
  P(\delta_i \mid h, d_i, s_i, k_i, \ell_i)
\end{multline*}

\noindent Each factor is modeled by a forest, with separate delay forests for client and server packets, yielding six forests in total. The packet-type forest only generates output for a server packet that opens the connection or follows a client packet, since a \emph{continuation} and client packets are always timed from the previous packet. $\ell_i$ holds the lengths of all records that begin in a packet, so the record forest draws one length at a time, in a loop, until the packet is covered. A packet filled entirely by a record that began in an earlier packet starts no new records.

We chose this order of feature generation based on the dependencies between packet features. Direction comes first because packet type and the \textsc{tail} size symbol depend on the sender. TLS record lengths follow packet size because \textsc{fill} depends on the space available in the packet. Delay comes last because its reference packet depends on the packet type, which also strongly influences the delay distribution. In our training data, the median delay is 5.3\,ms for replies but only 30\,µs for continuations.

\myparagraph{Training Forests}
Each of the six forests contains four trees, each grown on a different bootstrap sample of the training data~\cite{Breiman2001a} so that the trees partition it differently. A tree predicts the next value in a packet sequence. At each node, we choose the feature and threshold that minimize the entropy of the predicted value in the resulting partitions. For sufficiently long packet sequences, evaluating every feature at each node would be computationally expensive, so we randomly select $\lceil\sqrt{p}\rceil$ of the $p$ possible features~\cite{Liaw2002a,Probst2019a}. In addition, every node considers the packet position, features of the preceding packet, and features of the current packet that precede the predicted value in the generation order described above. For each candidate feature, we evaluate up to 16 thresholds at evenly spaced quantiles of its values among the training examples at that node~\cite{Chen2016a}. A tree stops growing when it reaches its maximum depth or a split would leave too few training examples in a child node. During generation, both endpoints use their shared secret to select the same tree. We report the tested hyperparameter values and the basis for each in~\Cref{tab:model-params} in the Appendix.

\myparagraph{Adapting to Personas} The forests are grown from flows of many servers, so sampling from their original counts would produce traffic representative of the broader TLS population. Instead, each proxy keeps the tree structure unchanged to preserve patterns learned across servers, but uses flows from its \persona{} to generate \schedule{}s specific to its selected server, which we call its \persona{}'s \textit{\profile{}}. To build each \profile{}, we pass the selected server's flows down every tree and count, at each node they reach, the value that came next, separately for each of the 20 packet positions, so that packet $i$ is drawn from what the \persona{}'s flows produced at packet $i$. Because \profile{} flows are few compared with the training flows (160 vs 8,000), many deep nodes may hold none of them. For a draw that finds no count at its node, we use the nearest node above it that has one. Since our architecture does not require training a new model of traffic, a proxy operator can give blocked servers \profile{}s built from different \persona{} servers with IP address rotation. Moreover, a \profile{} does not store exact flows of a server but a compact representation of them.

\subsection{Shaper Design and Implementation}
\label{sec:shaping}

\begin{figure*}[t]
\centering
 \includegraphics[width=\textwidth,keepaspectratio]{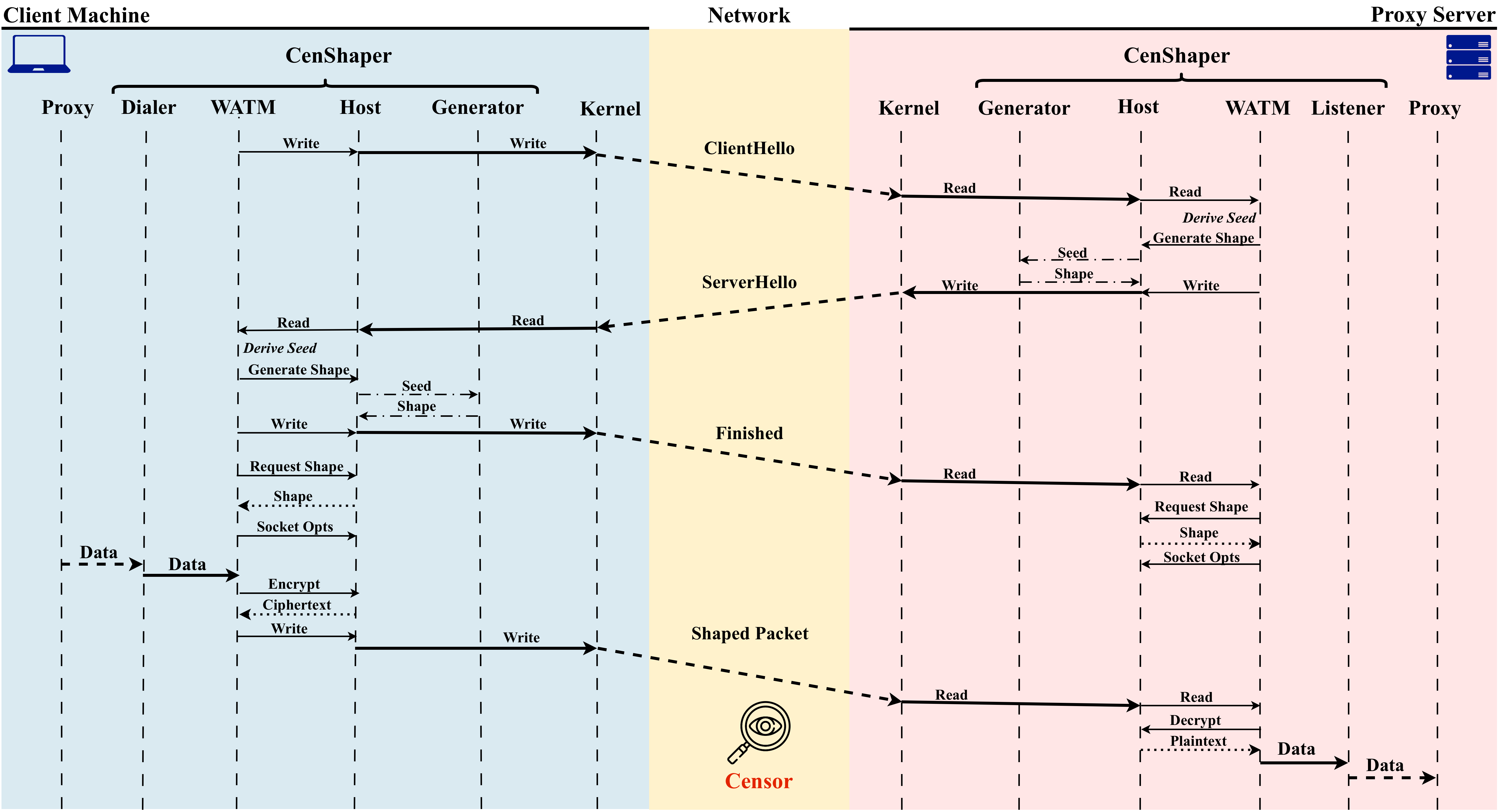}
\caption{\textbf{\censhaper{}'s Shaping Process.} Once the \texttt{ServerHello} is known, each end derives the connection's seed from the TLS handshake secret, and its host requests the shape from the Generator while the handshake continues. After the handshake, the WATM fetches the shape and sends the proxy's data in packets of the shape's sizes and delays, with the host encrypting each record. The figure shows one shaped packet; after the first 20, data is relayed unshaped.  Dashed lines are TCP connections, solid lines function calls, dotted lines return values, and dash-dotted lines pipes.}

\label{fig:censhaper}
\end{figure*}
We implemented \censhaper{} as a WATER transport module~\cite{Chi2024a}. WATER allows a host application to load network transports as WebAssembly modules at runtime. These modules can be updated without modifying the host application and run on any operating system the host supports. Our host runs alongside the proxy, similar to a Tor pluggable transport~\cite{TorPT}. To shape proxy traffic, \censhaper{} must control the outer TLS connection. Proxies such as Xray~\cite{XrayCore} and GOST~\cite{GOST} can be configured to disable their existing TLS layer, allowing \censhaper{} to provide TLS instead without modifying the proxies. This approach does not support protocols such as REALITY~\cite{xtlsreality}, which do not offer this configuration option.

\myparagraph{Module and Host}
The host provides the flow generator, socket control, and AES-GCM encryption. The generator runs as a separate process and holds the forest (5.7\,MB on disk) and the \persona{} \profile{}. Socket control allows \censhaper{} to read if TCP timestamps are in use, which changes the largest packet an endpoint can send (see \emph{Practical Adjustments}), and to set Linux's MSG\_EOR flag, which stops Linux from appending later data to a queued write. AES-GCM is done inside the host for efficiency: WebAssembly cannot use the processor's AES instructions, and the host encrypts a record that fills one packet about seven times faster than the module. The module performs the TLS1.3 handshake, using a fork of uTLS~\cite{utls} that mimics Chrome's \texttt{ClientHello} and can encrypt records of predetermined lengths. It also derives the seed using TLS handshakes and sends the proxy's data according to the shape it receives from the generator.

\myparagraph{Shape Synchronization} To generate the same shape, both ends need a shared seed (unknown to on path observer). \censhaper{} derives 32 bytes from the TLS~1.3 handshake secret by HKDF-Expand-Label~\cite{rfc8446}. Unlike the exporter's secret~\cite{rfc5705} used by \proxyshaper{} \cite{Ortwein2026a}, which is available only after the server's \texttt{Finished} message, the handshake secret is known to both ends as soon as the \texttt{ServerHello} is processed, so both can generate the shape during the handshake. The generator also rejects invalid \schedule{}s before they are transmitted.

\myparagraph{What to Send}
Each end keeps a buffer of unsent bytes per direction. For each of its packets, it seals the records that the shape says begin there, fills each with as much pending tunnel data as fits, and writes exactly the packet's size from the front of the buffer. A long record leaves in pieces over later packets, and several short records can share a packet, as they do from real servers. The rest of each record is TLS~1.3 padding~\cite{rfc8446} which gets removed by the receiver's TLS stack and is invisible to an observer. Shaping thus adds no message or field to the flow, but it requires TLS~1.3. 

\myparagraph{When to Send}
A packet leaves once its predetermined delay has passed since its reference packet. For the first packet, the client measures delay from when it sends out the Finished message of the handshake while the proxy measures it after receiving the message. Finally, we cap delays at 5\,s to prevent excessively long delays from holding back tunnel data.

\myparagraph{Recognizing Packet Boundaries}
The shaper sits above TCP, which delivers a stream of bytes, so it does not see packet boundaries. Since both ends know the shape of the flow in advance, each knows the byte offset at which the other's $i$-th packet ends, and the arrival of that byte counts as the arrival of the packet. 

\myparagraph{The End of the Window}
We consider a window complete only if each sender has sent out its own packets and received the other end's last one. A close requested during the window is held until the window is complete, because a window cut short is itself recognizable. A record left unfinished by the window's last packet leaves with the first data packet after the window, or with the TLS close message if the flow ends first.

\myparagraph{Practical Adjustments}
To allow the proxy header to travel in the first packet, the client waits up to 1\,ms for the application's first bytes. Moreover, some proxy protocols, such as Trojan, refuse headers that arrive in pieces. For such protocols, the first bytes of the flows are sealed into records that can carry at least 64 bytes and any smaller records before them only carry padding. The first adjustment affects delay slightly for the first packet while the second does not impact the shape of the flow in any capacity.

Moreover, on connections that use timestamps, a packet of size 1460 bytes will be split into two packets of 1448 bytes and 12 bytes. In such cases, we cut every generated packet size down to 1,448 bytes and shorten the record that covers its last byte by the same amount. We also disable session tickets and automatic record sizing, so that the TLS stack does not add records of its own.
\section{Evaluation}
\label{sec:eval}

We first evaluate whether \censhaper{} generates \schedule{}s consistent with individual TLS servers and whether classifiers trained on known proxies can identify those using unseen personas. We then test \censhaper{} with five proxy protocols, measuring detection by known traffic-analysis attacks, divergence from ordinary TLS, and implementation overhead.

\subsection{Generator Evaluation}
\label{sec:model-eval}

We build and evaluate \censhaper{}'s generator using only the TLS~1.3 flows of our capture, as \censhaper{} relies on its zero-padding. We randomly divide the flows into five sets for growing the trees (30\%), building \profile{}s (10\%), training classifiers (25\%), calibrating detection thresholds (10\%), and testing (25\%).

\myparagraph{Training Data} We train the forests using 8,000 randomly selected flows from the training set, excluding traffic from servers sharing an SNI or IP address with any of the 355 servers selected for building \profile{}. Training on larger samples may improve performance (Appendix~\ref{app:model-params}), but we found 8,000 flows, combined with a persona \profile{}, were sufficient to build a forest that generates \schedule{}s consistent with real flows of the selected server (\Cref{sec:generator-results}).

\myparagraph{Selecting Personas} We select servers with at least 320 flows, allowing us to construct \profile{}s from 160 flows and evaluate their generated \schedule{} against real held-out flows from the same server. We also exclude servers with traffic gaps exceeding 10\,s, since reproducing such delays would introduce excessive latency. This leaves 355 servers, of which we reserve 60 for model hyperparameter tuning (Appendix~\ref{app:model-params}) and use the remaining 295 (177 SNIs at 239 IP addresses) for evaluation.

\myparagraph{Classifiers}
We evaluate four classifiers from prior work on traffic classification and proxy detection. We reproduce their published implementations and training procedures as closely as possible, adapting input representations or architectural parameters where necessary for our 20-packet observation window.

\begin{itemize}
    \item \textbf{Deep Fingerprinting (DF)}~\cite{Sirinam2018a} Following Wails et al.~\cite{Wails2024a,Wails2025a}, we use signed packet sizes to encode size and direction. We retain the published training settings and architecture, except for kernel and pooling sizes, which we tune to accommodate our 20-packet inputs.
    
    \item \textbf{p-FP}~\cite{Oh2019a} We use signed TLS record lengths, one of the representations evaluated by Oh et al. We retain their CNN architecture and use HyperOpt's TPE to tune filter and pooling sizes for inputs of 40 record lengths.
    
    \item \textbf{Stacked Denoising Autoencoder (SDAE)}~\cite{Rimmer2018a} Following Wails et al.~\cite{Wails2024a}, we use signed packet sizes without modifying the original architecture.

    \item \textbf{eXtreme Gradient Boosting (XGBoost)}~\cite{Chen2016a} Tree-based classifiers have been used to detect proxy traffic~\cite{Ortwein2026a,Barradas2018a,Kamali2026a}, and their modest memory and computational requirements make them plausible for censors. Unlike the CNNs above, XGBoost supports our full feature set without architectural changes. We evaluate it using packet sizes and directions alone, as in~\cite{Ortwein2026a}, and with delays and TLS record lengths added.
    
\end{itemize}

\myparagraph{Transferability Settings} A classifier is useful for identifying proxies whose addresses are unknown, otherwise a censor can directly block them. To evaluate whether fingerprints learned from known proxies transfer to others (G2), we consider three settings:
\begin{enumerate}[label=(\roman*)]
  \item \emph{Seen}: The classifier is trained on \schedule{}s from every test \profile{}.
  \item \emph{Unseen-Server}: The classifier has not seen \schedule{}s from the test \profile{}, but may have seen \schedule{} from other \profile{}s sharing its SNI.
  \item \emph{Unseen-Service}: The classifier has not seen \schedule{}s from the test \profile{} or any other \profile{} sharing its SNI. This distinction matters because servers of the same service tend to exhibit similar traffic patterns (\Cref{fig:map}).
\end{enumerate}
For the first setting, we use 150,000 held-out flows per class for evaluation while using five-fold cross validation (30,000 per class, per fold) for each of the \textit{unseen} settings. The detection threshold is calibrated using the shapes of additional 30,000 real flows excluded from training, validation, and testing. We report the AUC and true positive rate at a 1\% false positive rate for evaluation.

\subsection{Results} 
\label{sec:generator-results}

\myparagraph{Profile Fidelity}
\Cref{sec:analysis-persona} showed that 160 recorded flows can represent a server's traffic. Here, we test whether our generator produces \schedule{}s consistent with that traffic. For each of 295 servers, we build \profile{}s using 20, 40, 80, 160, and 320 flows and generate 300 \schedule{}s per \profile{}. We compute $H$ between these \schedule{}s and 300 held-out flows from the corresponding server, excluded from \profile{} construction. The median $H$ decreases from 0.56\% with 20 flows to 0.27\% with 80, then changes little at 160 (0.26\%) and 320 (0.24\%) flows (\Cref{tab:persona-size}). These distances are small compared with the median 29\% between randomly selected servers, indicating that generated \schedule{}s follow their target servers' traffic distributions (G1). The diminishing improvements beyond 80 flows further support our conservative choice of 160 flows per \persona{}. Since the forests were trained without flows from these servers or others sharing their SNI or IP address, the results also show that the shared model can be adapted to servers absent from its training data.
\begin{table}[!t]
  \centering
  \setlength{\tabcolsep}{5pt} 
  \begin{tabular}{@{}l r r r r r@{}}
    \toprule
    Flows ($n$) & 20 & 40 & 80 & 160 & 320 \\
    \midrule
    File Size & 65\,KB & 91\,KB & 125\,KB & 173\,KB & 230\,KB \\
    Energy Coefficient ($H$) & 0.56\% & 0.36\% & 0.27\% & 0.26\% & 0.24\% \\
    \bottomrule
  \end{tabular}
  \caption{Median file size and $H$ (to held-out flows) of 295 evaluated servers for $n$ number of flows. After 80 flows we see very little reduction in $H$ which is consistent with our conservative threshold selection for persona size.}
  \label{tab:persona-size}
\end{table}


\myparagraph{Classifier Results}
\Cref{tab:generator-results} shows that classifiers evaluated on \schedule{}s generated using unseen services perform near random guessing ($\approx$ 0.5 AUC). At a 1\% false-positive rate, the best-performing classifier (highest AUC) detects only 8.5\% of traffic shapes generated using \profile{}s of unseen TLS services. The detection rises to 29\% when the classifier has seen the TLS service at a different IP address and 56\% for a classifier that has been trained on every \profile{}. The AUC, similarly, increases from 0.59 to 0.79 and then to 0.97. These results suggest that diversity of \profile{} built from different TLS servers limits what a censor can reuse across proxies (G2). A censor that observes less than 20 packets performs even worse (Appendix~\Cref{tab:classifier-pkt-count}).
\begin{table}[t]
  \centering
  
  \begin{tabular}{@{}l r r r r r r@{}}
    \toprule
    & \multicolumn{2}{c}{Seen} & \multicolumn{2}{c}{Unseen Server} & \multicolumn{2}{c}{Unseen Service} \\
    Classifier & AUC & TPR & AUC & TPR & AUC & TPR \\
    \midrule
    XGBoost & 0.94 & 47\% & 0.74 & 27\% & 0.55 & 11.2\% \\
    \quad all features & 0.97 & 56\% & 0.79 & 29\% & 0.59 & 8.5\% \\
    DF & 0.82 & 13\% & 0.64 & 6.6\% & 0.44 & 2.1\% \\
    p-FP & 0.67 & 4.5\% & 0.58 & 1.5\% & 0.46 & 0.8\% \\
    SDAE & 0.61 & 3.0\% & 0.57 & 2.1\% & 0.48 & 0.5\% \\
    \bottomrule
  \end{tabular}
  \caption{Classifiers trained against real traffic and traffic schedules generated using 295 personas (177 SNIs). TPR shows how many generated traffic schedules were correctly flagged by the classifier with a FPR of 1\%. An AUC of 0.5 represents classifiers performing near chance.}
  \label{tab:generator-results}
\end{table}

\subsection{\censhaper{} on the Wire}
\label{sec:on-wire}
To test \censhaper{} end-to-end, we run a client and proxy server with \censhaper{} on two x86-64 Linux machines with a handshake RTT of 124\,ms. An automated Firefox client, driven by Playwright, visits 2,000 randomly sampled domains from the top 10,000 in the Chrome UX Report~\cite{CrUX}, one at a time, through five Xray protocols: VLESS, VMess, Trojan, Shadowsocks, and VLESS over WebSocket (VLESS+WS). These visits produced a total of 146,263 flows. Although a deployed proxy will use a single \persona{}'s \profile{}, we evaluate \profile{}s built from 100 \persona{}s by loading them into the generators at both endpoints before the first connection and selecting one per connection using its shared secret. On average, each \profile{} was used for approximately 1,460 flows.

We evaluate XGBoost, the best-performing classifier from \Cref{sec:generator-results}, on real and shaped traffic for each of the three settings defined in \Cref{sec:model-eval}. To isolate the timing patterns introduced by \censhaper{}, we measure packet delays from their reference packets to when the shaper hands them to the TCP stack. This excludes TCP congestion control~\cite{rfc6928} and the testbed's longer round-trip time, which could otherwise allow the classifier to distinguish the testbed's network conditions from those of the ISP traffic rather than detect artifacts of the shaper. We evaluate timing precisions of 1, 10, and 50\,ms, as well as a classifier without timing features (\Cref{tab:on-wire-classification}). Even at 1\,ms precision and a 1\% false positive rate, the classifier detects only 8.2\% of flows shaped using \profile{}s of unseen TLS services. As in \Cref{tab:generator-results}, detection increases when the classifier has seen other \profile{}s of the same service (42\%) or the test \profile{} itself (77\%).

\begin{table}[t]
  \centering
  \begin{tabular}{@{}l r r r r r r@{}}
    \toprule
    & \multicolumn{2}{c}{Seen} & \multicolumn{2}{c}{Unseen Server} & \multicolumn{2}{c}{Unseen Service} \\
    Delay precision & AUC & TPR & AUC & TPR & AUC & TPR \\
    \midrule
    1\,ms     & 0.99 & 77\%   & 0.91 & 42\% & 0.77 & 8.2\% \\
    10\,ms    & 0.98 & 60\%   & 0.85 & 27\% & 0.68 & 4.0\% \\
    50\,ms    & 0.98 & 58\%   & 0.86 & 26\% & 0.68 & 4.1\% \\
    No delays & 0.98 & 56\%   & 0.86 & 27\% & 0.69 & 3.2\% \\
    \bottomrule
  \end{tabular}
  \caption{XGBoost trained on real and shaped flows, with delays given at decreasing precision. TPR (Detection) is reported at 1\% FPR (Collateral).}
  \label{tab:on-wire-classification}
\end{table}

\subsection{TLS-over-TLS Fingerprinting}

Xue et al.~\cite{Xue2024a} demonstrated that proxies tunneling user TLS traffic through TLS-encrypted connections can still be detected from packet sizes and timing indicative of encapsulated TLS handshakes. Their method uses $\chi^2$ and Mahalanobis-distance tests trained to distinguish TLS from non-TLS traffic using the first 25 data packets. After removing the outer TLS handshake, they flag connections whose remaining traffic is still classified as TLS, indicating an encapsulated handshake. We evaluate whether \censhaper{} hides this fingerprint in traffic captured on the wire. We implement the same tests but tune them using unshaped proxy traffic instead of non-TLS traffic, allowing the classifiers to learn differences in proxy traffic and normal TLS traffic. Since we are directly comparing traffic from \censhaper{} against proxy traffic and normal TLS traffic, we do not strip the outer handshake. 

To collect baseline proxy data, we repeat the experiment described in \Cref{sec:on-wire} without \censhaper{}. Any connections with fewer than 25 packets is excluded which leaves a total of 113,438 unshaped proxy flows. We sample an equal number of TLS~1.3 flows from our ISP capture as all our proxies use TLS~1.3 for the outer tunnel. Evaluation is done using five-fold cross-validation, allocating two folds to classifier training, two to threshold calibration, and one to testing in each round. The trained classifiers are then evaluated on the shaped flows from \Cref{sec:on-wire}—subject to the same 25-packet minimum—to compare detection rates between shaped and unshaped proxy traffic.

\Cref{tab:tls-over-tls} reports detection rates for shaped and unshaped proxy traffic at false positive rates of 1\% and 5\% on ordinary TLS flows. At a 1\% false positive rate, the classifier only detects $\approx$3\% of the flows shaped by \censhaper{} while correctly flagging 85\% of unshaped proxy flows. Even at a 5\% false positive rate, detection remains below 17\% for shaped traffic while exceeding 97\% for unshaped traffic. 

\begin{table}[t]
  \centering
  \setlength{\tabcolsep}{3.5pt} 
  \begin{tabular}{@{}l r r r r r r@{}}
    \toprule
    & \multicolumn{3}{c}{Unshaped} & \multicolumn{3}{c}{Shaped} \\
    \cmidrule(lr){2-4}\cmidrule(l){5-7}
    Protocol & Flows & 1\% FPR & 5\% FPR & Flows & 1\% FPR & 5\% FPR \\
    \midrule
    VMess       & 24,206 & 89.1\% & 97.9\% & 28,640 & 3.1\% & 16.6\% \\
    VLESS       & 19,988 & 87.7\% & 97.6\% & 28,624 & 2.9\% & 16.7\% \\
    VLESS+WS    & 24,153 & 85.6\% & 97.1\% & 29,305 & 2.9\% & 16.7\% \\
    Trojan      & 22,449 & 88.8\% & 97.9\% & 29,410 & 3.0\% & 16.5\% \\
    Shadowsocks & 22,642 & 89.7\% & 97.9\% & 30,283 & 2.9\% & 16.4\% \\
    \bottomrule
  \end{tabular}
  \caption{Percentage of proxy flows detected at 1\% and 5\% collateral damage using classifiers adapted from Xue et al.~\cite{Xue2024a}. Flows are evaluated using five-fold cross-validation.}
  \label{tab:tls-over-tls}
\end{table}

\subsection{Similarity to Real-World TLS}

Our classifier experiments demonstrate that \censhaper{} reduces detection by specific attacks, but do not establish how similar shaped traffic is to packet timing, directions, and sizes of real TLS. We therefore use TLS Divergence~\cite{Ortwein2026a}, an approximation of KL divergence, to compare on-wire proxy traffic with real TLS traffic. For each protocol, we compare shaped and unshaped proxy traffic separately against 150,000 TLS~1.3 flows sampled from our ISP dataset. Across all five protocols, our on-wire datasets contain 113,439 unshaped flows and 146,263 shaped flows. We analyze each flow's first 20 post-handshake data packets.

\Cref{tab:tls-div} shows the divergence for each protocol. Unshaped traffic has a divergence of 14.4 to 19.7 nats, with VLESS+WS exhibiting the highest divergence. Its WebSocket upgrade, also identified by Ortwein et al.~\cite{Ortwein2026a} as a major source of early-packet divergence, consists of a deterministic 618-byte request followed by a 151-byte response in our setup. The first three packets contribute 5.5 nats of divergence, compared with 3.4 nats for the other protocols. With shaping, divergence falls to $\approx$1.9 nats across all five protocols, a reduction of approximately sevenfold or more. 
\begin{table}[t]
\centering
\footnotesize
\setlength{\tabcolsep}{4pt}
\begin{tabular}{lccccc}
\toprule
 & VMess & VLESS & VLESS/WS & Trojan & Shadowsocks \\
\midrule
Unshaped  & 14.4  & 14.8  & 19.7 & 14.9  & 14.9  \\
Shaped    & 1.9  & 1.92  & 1.90  & 1.90  & 1.93  \\
Reduction & 7.6$\times$ & 7.7$\times$ & \textbf{10.4$\times$ }& 7.8$\times$ & 7.72$\times$ \\
\bottomrule
\end{tabular}
\caption{TLS Divergence (nats)~\cite{Ortwein2026a} of proxy flows from real TLS~1.3 flows before and after shaping.} 

\label{tab:tls-div}
\end{table}


\subsection{Performance}
\label{sec:performance}
\censhaper{}'s base model ships in less than 6 MB and a \persona{}'s \profile{} in a median of just 173 KB. On our client machine, loading a single \profile{} took about 0.3\,s. The memory peaked at 112\,MB, with a stable value of 62\,MB after loading. Generating a single traffic shape takes about 0.25\,ms on a single core, well within the time duration of handshake between client and proxy server. One remaining source of observable overhead is the TLS handshake, which runs in WebAssembly. The client sent its \texttt{ClientHello} 10.8\,ms after the TCP handshake and its \texttt{Finished} message 10.1\,ms after the server's reply arrived, against 1.3\,ms and 1.0\,ms for Xray's own TLS on the same path. The proxy answered a \texttt{ClientHello} after 8.9\,ms, against 1.7\,ms without shaping. This can be fixed by moving the outer handshake's public-key operations to the host, as we did for AES-GCM (\Cref{sec:shaping}).

\begin{figure}[t!]
\centering
 \includegraphics[width=\columnwidth,keepaspectratio]{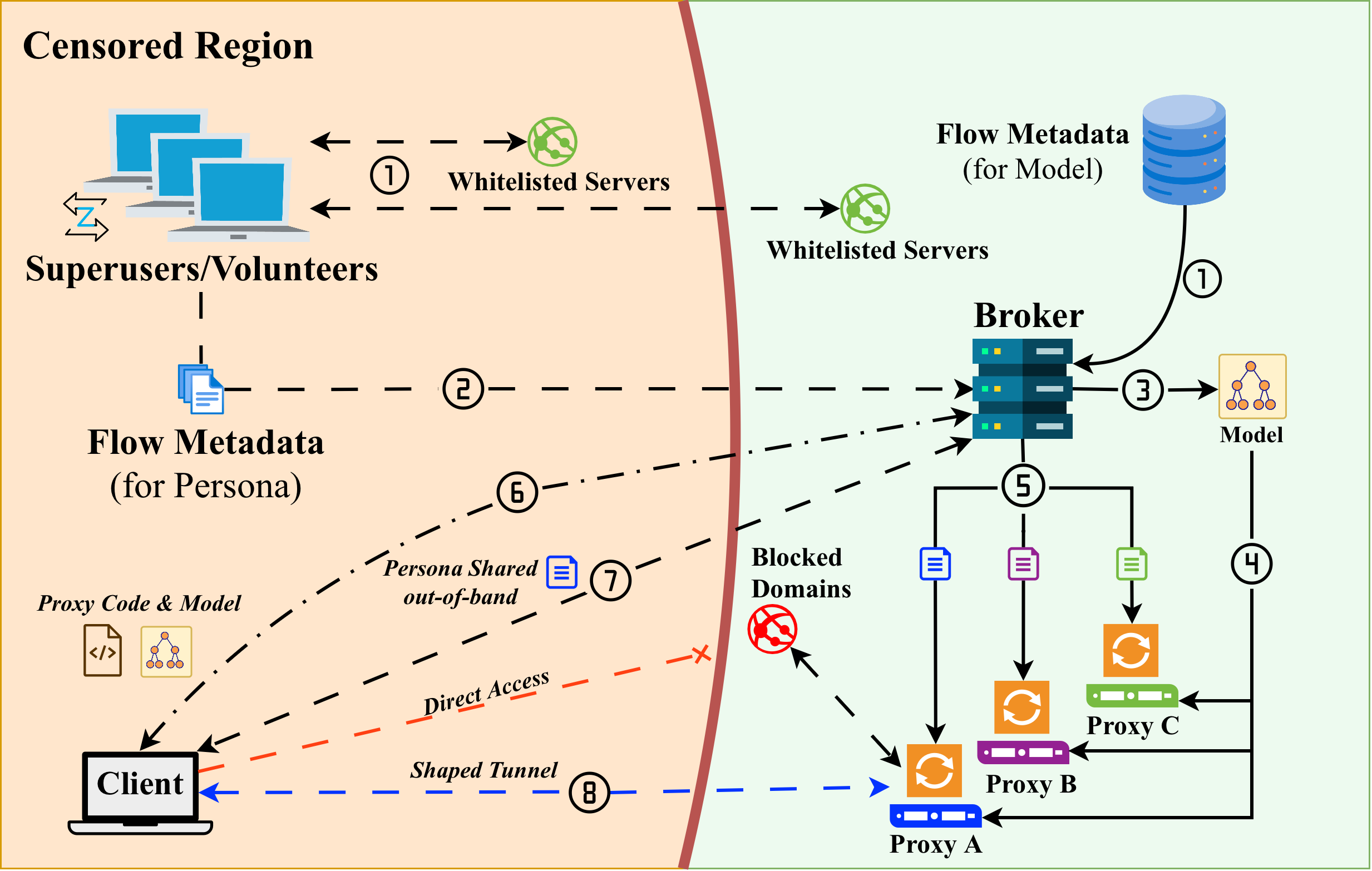}

\caption{\textbf{Deploying \censhaper{}.} \fstep{1} Flow metadata for the base model and \persona{}s is collected by volunteers. \fstep{2} The metadata is sent to the central broker, which \fstep{3} trains the model, \fstep{4} sends it to the proxies, and \fstep{5} assigns each proxy a different \persona{}. \fstep{6} Clients receive the model with the proxy code and \fstep{7} their proxy's \persona{} out-of-band. \fstep{8} The tunnel connection is then shaped according to the \persona{}.}
\label{fig:workflow}
\end{figure}
\section{Deployment Considerations}
\label{sec:deployment}

\Cref{fig:workflow} shows how \censhaper{} could be integrated into circumvention tools, based on our analysis and discussions with developers whose tools serve millions of users~\cite{lantern, Snowflake2023Metrics}. We describe each stage of the pipeline below:

\myparagraph{What to Collect}
\censhaper{} requires relatively little data. In our evaluation, 8,000 flows were sufficient to train the base model, and 160 flows from its selected \persona{} to build \profile{}. Although these sample sizes do not guarantee resistance to every censor, our evaluation shows that they suffice to generate traffic that more closely matches ordinary TLS and evade the classifiers evaluated in \Cref{sec:eval}. We collect only packet and TLS metadata, along with the information needed to identify individual servers, without recording packet payloads or client IP addresses. We release the Zeek script used to extract this information at the point of capture (\Cref{fig:workflow} \fstep{1}--\fstep{2}). 

We recommend selecting \persona{}s that censors cannot afford to block. Some censors make this set explicit. During mobile internet shutdowns, Russia keeps a list of sites that remain reachable, which its Digital Ministry compiles from the most popular resources in the country and updates weekly~\cite{Interfax2025a}. Recording only flows to such allowed servers also limits what a recording can reveal about its contributor.

\myparagraph{Where to Collect}
A key takeaway from \Cref{sec:model-eval} is that each proxy should imitate a different server to prevent a censor's knowledge from scaling. Since \profile{}s can be built independently of the base model, the model can be trained once and new \profile{}s added later, potentially using flows recorded elsewhere. However, \profile{}s recorded outside a region may share artifacts of the network from which they were recorded, and a censor that learned such artifacts could potentially block multiple \profile{}s at once. Therefore, we recommend that at least the \profile{}s be recorded inside the region where they are used. \Cref{fig:workflow} therefore shows two sources of flow metadata: flows for the model, which may come from any network, and flows for \profile{}s, which volunteers record inside the region (\fstep{1}) and send to the broker (\fstep{2}).

\myparagraph{Who Collects Data}
Existing measurement platforms already operate through volunteers inside censored regions. In October 2026, volunteers submitted OONI Web Connectivity measurements from 48 networks in Iran, 19 in China and 556 in Russia~\cite{OONI_API}. Each of these measurements opens connections to the websites of a test list, and OONI's probes already record the size and time of every read and write on those connections~\cite{OONINetEvents}. Recording the same at the level of packets and TLS records would make these connections a source of \profile{}s and help provide useful data to counter traffic analysis. Such requests have precedent: OONI started monitoring STUN servers on request from the community after Google's STUN server was blocked in China~\cite{OONISTUN}. Alternatively, volunteers could collect and contribute traffic metadata independently of existing measurement platforms, but would need administrator rights to capture traffic on desktop devices or root access on mobile devices.

\myparagraph{Building Model \& Persona}
Training and distribution can be handled by a central broker, such as the configuration service of a circumvention app, from which Lantern's and Psiphon's clients fetch their servers~\cite{LanternRadiance,PsiphonRSL}; a bridge distributor, such as Tor's rdsys~\cite{Rdsys2025}; or the subscription server of a commercial proxy service~\cite{Habib2026a}. The broker trains the base model (\Cref{sec:model}, \fstep{3}) and builds a \profile{} for each server that meets the following criteria: it supports TLS 1.3, its \profile{} matches held-out traffic, its traffic distribution is relatively stable over time (\Cref{tab:ttl}), and its sending delays are not excessively long. The broker distributes the shared model to proxy servers (\fstep{4}), and clients receive it with the proxy code (\fstep{6}). Each proxy receives its own \profile{} (\fstep{5}), and no two proxies receive \profile{}s built from the same service. A client obtains its proxy's \profile{} and IP address through the same out-of-band channel (\fstep7). The broker must therefore keep the \profile{}s, the flows they are built from, and their assignment to proxies as confidential as the proxies' addresses; otherwise a censor could generate the traffic shapes of undiscovered proxies and train a classifier against each, undermining the benefits of compartmentalization (G2). 

\myparagraph{Communication} Once the client and proxy have the model and \profile{}, the client can reach blocked domains through the shaped tunnel (\fstep{8}). Since \profile{}s are built using stable \persona{}, the broker only needs to rebuild them when they no longer match fresh traffic from their servers. If a proxy is blocked, it can be redeployed with a new IP address and \profile{} without retraining the model or affecting other proxies. A proxy server can also use different \profile{}s depending upon the region of the clients, as the allowed domains may vary between regions.

\section{Limitations}

A key contribution of this work is grounding traffic shaping in measurements of millions of real-world TLS connections. However, our dataset comes from a single regional U.S. ISP, and we lack comparable measurements from networks in censored regions. The dataset does, however, include globally popular services, such as Akamai, Google, Tencent, TikTok, and Apple (\Cref{fig:map}). More, we find that different IP addresses hosting the same service exhibit similar traffic shapes. While the ideal approach is to record \persona{} \profile{}s in the regions where they are used (\Cref{sec:deployment}), the fact that a service's traffic shape is largely driven by the service itself gives us confidence in the design.

A censor could potentially create a mapping between traffic shapes of popular services and their IP addresses, allowing the censor to detect a proxy if there is mismatch between the shape of a flow and its corresponding IP address. But without knowing the selected \persona{}, which \censhaper{} requires to be kept confidential, to detect the mismatch, the censor must first identify the imitated service from observed traffic shapes. This means that the censor would need to collect ground-truth traffic from many TLS services, train a classifier to identify services from their traffic shapes, and keep its training data up-to-date as those services change. In contrast, the broker can select a new \persona{} and distribute its \profile{} without retraining the model. This asymmetry favors the proxy provider: rotating a \persona{} is cheap, while keeping up with every possible \persona{} is not.

\censhaper{} shapes the first 20 data packets of each connection. Censors have blocked circumvention tools based on the first data packet alone~\cite{Ablove2026a,Wu2023a}, and 20 packets gives us a wide margin beyond that. The window is not a limit of the design: providers can shape fewer packets with the same model and \persona{} \profile{}, or extend shaping to later traffic by retraining the model on longer flows, should censors begin analyzing it. Keeping the window short keeps shaping and the job of broker inexpensive.  Extending shaping to the entire connection would require selecting a \persona{} whose traffic shape matches the application's workload to avoid excessive performance degradation. While the broker could select \persona{}s based on expected workloads, adapting to individual applications would require closer integration with the proxy, sacrificing \censhaper{}'s ability to run without changes to existing proxies.

\section{Conclusion}
\label{sec:conclusion}
Traffic analysis is no longer a distant threat to circumvention, and providers need shaping defenses they can deploy today. \censhaper{} draws on real-world traffic to decide what shape proxy traffic should take. In our measurements, TLS services had distinct traffic shapes that stayed relatively stable over three days. \censhaper{} gives each proxy a distinct traffic shape drawn from a real TLS server. Across proxies, the resulting traffic is \emph{ordinary, but not identical}. \censhaper{} is built for how proxy providers actually operate. Creating a persona \profile{} takes a small recording rather than retraining, there is no need for coordination between client and proxy, and providers can deploy it without changing their proxies or workflow. Across five popular proxy protocols, it largely defeats TLS-over-TLS fingerprinting, and fingerprints learned from known proxies transfer poorly to proxies imitating unseen services. \censhaper{} does not make proxies invisible, but it changes what a censor must do: learn the service each proxy imitates, not one fingerprint for an entire provider. We hope it gives circumvention developers one more practical layer of defense, and researchers a foundation to build on.
\bibliographystyle{plain}
\bibliography{paper}
\appendices

\section{Ethics Considerations}
\label{sec:ethics}

Handling real user data, especially at the scale of an ISP, requires important ethical considerations. Although our IRB categorized this research project as ``Not Regulated,'' we still took extensive precautionary measures to ensure user anonymity and data privacy. We received formal approval from the ISP for the information collected during this project and followed their established guidelines for handling sensitive data. We also restricted our data collection and analysis to the minimum amount of information required for our analysis. We monitor only TCP port 443 and use Zeek's TCP and TLS analyzers to log, for every packet of a connection, its timestamp, direction, TCP flags, payload size, and sequence and acknowledgment numbers; for every TLS record, its content type (e.g., handshake or application data) and length; and for each connection, the negotiated TLS version, the server's address and its SNI. No payload is written to disk, and the client's address is not logged. The raw data was never moved from the network tap, and any analysis which required server's address and SNI was performed on the ISP operated machine solely. In our post-analysis stage, we move only the metadata of packets (direction, size of payload, packet delay, and lengths of TLS records) to a separate university-owned machine equipped with a GPU to evaluate our model of TLS traffic against advanced machine learning classifiers.
\begin{figure*}[t]
  \centering
  \includegraphics[width=\textwidth]{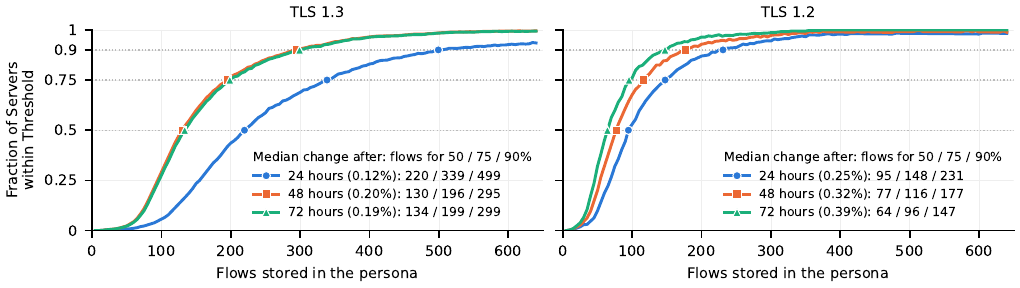}
  \caption{\textbf{Fraction of $n$ flow \persona{}s within threshold, all servers.} Evaluated over every server with at least $2n$ flows, so the set of servers shrinks as $n$ grows (TLS~1.3: 19,872 servers at $n=5$ and 4,200 at $n=160$). Thresholds and step size are the same as in \Cref{fig:persona-cdf}.}
  \label{fig:appendix-persona-cdf}
\end{figure*}

\section{Persona Budget on All Servers}
\label{app:persona-cdf}
\Cref{fig:appendix-persona-cdf} repeats the analysis of \S\ref{sec:analysis-persona} without the 1,000-flow minimum: at each persona size $n$, it includes every server with at least $2n$ flows. The curves keep the same shape and order, but more flows are needed. Half of TLS~1.3 servers meet the 24-hour threshold at 220 flows and 90\% at 499, compared with 156 and 441 for servers with at least 1,000 flows. For TLS~1.2, the corresponding values are 95 and 231. Because the set of servers shrinks as $n$ grows, from 19,872 TLS~1.3 servers at $n=5$ to 1,131 at $n=640$, each point is the fraction of servers tested at that size, not a distribution of the flows each server needs.

\section{Classifier Details}
\label{app:classifiers}

\myparagraph{Configurations}
\Cref{tab:censors} lists the configuration of each classifier in \S\ref{sec:model-eval}. We keep the published architectures and training settings wherever our 20-packet inputs allow, and tune only the settings that depend on input length: the kernel and pooling sizes of DF and the filter and pooling sizes of p-FP. For XGBoost, the original paper does not specify the exact hyperparameters tested~\cite{Ortwein2026a}, hence we select the depth, learning rate, and minimum leaf weight by grid search on validation AUC, and stop adding trees once validation AUC no longer improves.

\myparagraph{Fewer Packets}
\Cref{tab:classifier-pkt-count} evaluates classifiers using fewer packets from each connection. Detection falls with fewer packets in all three settings. With five packets, the classifier detects 27\% of \schedule{}s from seen \profile{}s and 2.3\% from unseen services, compared with 56\% and 8.5\% with 20 packets.

\section{Forest Parameters}
\label{app:model-params}
We select the generative forest's hyperparameters on the 60 servers reserved in \S\ref{sec:model-eval}, none of which is used for evaluation. For each alternative, we change one setting, regrow the trees if the setting affects them, build a \persona{} of 160 flows for each server, and compute $H$ between its generated flows and held-out real flows of the same server, with at most 1,024 flows on each side. We compare each alternative with our model using using the 60 server set aside for validation. \Cref{tab:model-params} shows that removing any component of the model makes generated flows significantly less similar to their server. Among the hyperparameters, maximum depth of 10 and Gini perform significantly worse. The other alternatives differ from our model by about as much as generating \schedule{}s using different seeds. The number of random candidate features and the sample of training flows for each tree follow the literature~\cite{Liaw2002a,Breiman2001a} and were not varied.
\begin{table}[t]
\centering
\setlength{\tabcolsep}{4pt}
\begin{tabular}{@{}l l r@{}}
\toprule
Setting (our value) & Alternative & $H$ \\
\midrule
Our model & & 0.29\% \\
\quad other generation seeds & & 0.24\% \\
\midrule
\multicolumn{3}{@{}l}{\textit{Hyperparameters}} \\
Maximum depth (14) & 10 & 0.36\% \\
 & 18 & 0.22\% \\
Minimum examples per child & 8; 4 & 0.28\% \\
\quad (16; 8 for delays) & 16; 16 & 0.24\% \\
 & 32; 16 & 0.26\% \\
 & 128; 64 & 0.33\% \\
Trees per forest (4) & 2 & 0.27\% \\
 & 8 & 0.24\% \\
 & 16 & 0.26\% \\
Thresholds per feature (16) & 8 & 0.24\% \\
Split criterion (entropy) & Gini & 0.39\% \\
Training flows (8,000) & 32,000 & 0.25\% \\
\midrule
\multicolumn{3}{@{}l}{\textit{Ablation Runs}} \\
Always-considered features & drawn at random & 0.85\% \\
 & same, $2\lceil\sqrt{p}\rceil$ & 0.37\% \\
Per-position counts & pooled & 0.81\% \\
\textsc{tail} & sizes in bytes & 0.58\% \\
\textsc{fill} & lengths in bytes & 0.57\% \\
\bottomrule
\end{tabular}
\caption{Effect of changing one setting of the model at a time: median $H$ between generated and held-out flows over the 60 selection servers.}
\label{tab:model-params}
\end{table}

\begin{table}[t]
  \centering
  \begin{tabular}{@{}l r r r r r r@{}}
    \toprule
    & \multicolumn{2}{c}{Seen} & \multicolumn{2}{c}{Unseen Server} & \multicolumn{2}{c}{Unseen Service} \\
    Packet Count & AUC & TPR & AUC & TPR & AUC & TPR \\
    \midrule
    5  & 0.91 & 27\% & 0.66 & 11\% & 0.40 & 2.3\% \\
    10 & 0.94 & 40\% & 0.71 & 17\% & 0.47 & 3.4\% \\
    15 & 0.96 & 49\% & 0.75 & 24\% & 0.55 & 6.1\% \\
    20 & 0.97 & 56\% & 0.79 & 29\% & 0.59 & 8.5\% \\
    \bottomrule
  \end{tabular}
  \caption{XGBoost on all features, given only the first $n$ data packets of each flow and the TLS records that begin in them: AUC, and share of generated flows detected (true-positive rate) at a 1\% false-positive rate.}
  \label{tab:classifier-pkt-count}
\end{table}

\begin{table*}[t]
\centering
\begin{tabularx}{\textwidth}{@{}>{\raggedright\arraybackslash}p{2.6cm} >{\raggedright\arraybackslash}p{7.4cm} >{\raggedright\arraybackslash}X@{}}
\toprule
Setting & Value & Basis \\
\midrule
\multicolumn{3}{@{}l}{\textit{XGBoost}~\cite{Chen2016a}} \\
Input & directions and sizes; or directions, sizes, delay buckets and 40 record lengths & directions and sizes as Ortwein et al.~\cite{Ortwein2026a}; delays and record lengths added by us \\
Depth; rate; leaf weight & 9; 0.2; 1 & grid of $\{3,5,7,9\}\times\{0.03,0.08,0.2\}\times\{1,3,10\}$, highest validation AUC with all features; reused for directions and sizes \\
Trees & set per task; 1,233--1,995 in \Cref{tab:generator-results} & early stopping: at most 2,000, stop after 50 without a higher validation AUC \\
Other & logistic loss, L2 weight 1, histogram splits & logistic loss for two classes; the others are XGBoost defaults \\
\midrule
\multicolumn{3}{@{}l}{\textit{DF}~\cite{Sirinam2018a}} \\
Input & 20 signed packet sizes & Wails et al.~\cite{Wails2024a} \\
Layers & 4 blocks of 2 convolutions (32 to 256 filters), batch normalization, ELU then ReLU, max pooling, dropout 0.1; 2 dense layers of 512, dropout 0.7 and 0.5; 2 outputs & Sirinam et al. \\
Kernel; pool; stride & 3; 2; 2 & kernel and pool in $\{2,3\}$ with stride 2, the sizes whose windows fit 20 packets; Sirinam et al.\ searched 2--16 for 5,000 packets \\
Training & Adamax, rate 0.002, batch 128, 30 epochs & Sirinam et al. \\
\midrule
\multicolumn{3}{@{}l}{\textit{p-FP}~\cite{Oh2019a}} \\
Input & up to 40 signed TLS record lengths, zero-padded & Oh et al.'s record representation \\
Layers & 2 convolutions of 128 filters, tanh, max pooling, local response normalization; dense layer of 256, dropout 0.2; 2 outputs & Oh et al.'s table; their code uses ReLU and a keep probability of 0.8 \\
Filter; pool & 14; 3 & TPE, 30 trials, filter up to $\min(16,\lceil 40/\text{pool}\rceil)$, pool 2--6; Oh et al.\ searched filter 2--16, pool 2--50 \\
Training & SGD, rate 0.05 decayed by 0.96 per 1,000 steps, batch 30, L2 weight 0.001 on convolutions & Oh et al.; decay and L2 from their code \\
Epochs & at most 50, stop after 3 without a gain of 0.001 in validation AUC, keep the best & Our Setting \\
\midrule
\multicolumn{3}{@{}l}{\textit{SDAE}~\cite{Rimmer2018a}} \\
Input & 20 signed packet sizes & Wails et al.~\cite{Wails2024a} \\
Layers & 3 tanh layers of 1,000, 500, 300, dropout 0.1; 2 outputs & Rimmer et al. \\
Pretraining & per layer: 10\% of inputs set to zero, linear decoder, squared error; SGD, rate 0.1; 50, 10, 10 epochs, batch 128 & Rimmer et al.; epochs and batch from their example configuration \\
Training & SGD, rate 0.001, Nesterov momentum 0.9, batch 32, 30 epochs & Rimmer et al.\ and their configuration \\
\bottomrule
\end{tabularx}
\caption{Settings of the classifiers and the basis of each value. Searched settings were chosen once, on the validation flows of the seen setting, and kept for every fold. Before training neural networks, packet sizes are divided by 1,460 and record lengths by 16,406.}
\label{tab:censors}
\end{table*}

\end{document}